\documentclass[12pt, prd, showpacs]{revtex4}

\usepackage{amsmath}
\usepackage{amssymb}

\begin{document}

\title{The angular momentum and mass formulas for rotating stationary
quasi-black holes}
\author{Jos\'{e} P. S. Lemos}
\affiliation{Centro Multidisciplinar de Astrof\'{\i}sica, CENTRA,
Departamento de F\'{\i}%
sica, Instituto Superior T\'ecnico - IST, Universidade T\'{e}cnica de Lisboa
- UTL, Av. Rovisco Pais 1, 1049-001 Lisboa, Portugal\,\,}
\email{lemos@fisica.ist.utl.pt}
\author{Oleg B. Zaslavskii}
\affiliation{Astronomical Institute of Kharkov V.N. Karazin National
University, 35
Sumskaya St., Kharkov, 61022, Ukraine}
\email{ozaslav@kharkov.ua}

\begin{abstract}
We consider the quasi-black hole limit of a stationary body when its
boundary approaches its own gravitational radius, i.e., its
quasi-horizon.  It is shown that there exists a perfect correspondence
between the different mass contributions and the mass formula for
quasi-black and black holes in spite of difference in derivation and
meaning of the formulas in both cases.  For extremal quasi-black holes
the finite surface stresses give zero contribution to the total
mass. Conclusions similar to those for the properties of mass are
derived for the angular momentum.
\end{abstract}

\maketitle




\section{Introduction}

In \cite{paper1} we found the mass formula for static quasi-black holes.
There, we have defined a quasi-black hole as the limiting configuration of a
body, either non-extremal or extremal, when its boundary approaches the
body's own gravitational radius. This definition is an enlargement from a
previous definition \cite{qbh} (see also
\cite{mim,extdust,bon-d,shell,glue1}
), which applied to configurations with extremal matter (i.e., with mass
density equal to charge density in appropriate units) and spherically
symmetric, to a generic static case with no particular symmetry neither some
specific matter. In particular, the conclusion made in \cite{qbh}, was that
static quasi-black holes, where matter has finite surface stresses, should
be extremal, where the condition of finiteness of stresses is important for
the conclusion. Then, in \cite{paper1} the appearance of infinite stresses
was allowed, and the condition of extremal matter could be dropped, i.e.,
non-extremal matter was also admitted. Notably, the mass formula found in
\cite{paper1}, also implies, when the derivation of the formula is taken
into account, that for finiteness stresses, quasi-black holes must be
extremal. For the static extremal case, the existence of a quasi-black hole
requires electric charge, or some other form of repulsive matter, such as in
gravitational monopoles \cite{luewei1,luewei2}. Although, the extremal
condition in the electrical case may be achieved if a tiny fraction, 10$
^{-18}$, of neutral hydrogen loses its electron, the requirement of charge
somewhat bounds the astrophysical significance of such quasi-black holes.
Dropping the extremal condition for the matter, infinite stresses appear at
the quasi-black hole threshold, which makes these non-extremal objects quite
unphysical, although as argued in \cite{paper1}, consideration of such
systems has at least a systematic interest since it helps to understand
better the distinction between non-extremal and extremal limits, and the
relationship between quasi-black holes and black holes. The related issue of
gravitational collapse to a quasi-black hole state, always an important
problem, was treated preliminary in \cite{visser}.

Now, the rotational counterpart of quasi-black holes were found first by
Bardeen and Wagoner back in 1971 \cite{bardwag}. They discussed rotating
thin disks and found that for rotation less than extremal, the exterior
metric does not yield a Kerr vacuum spacetime, but for extremal rotation
(i.e, mass equal to angular momentum per unit mass) of the disk, and in this
case only, the exterior metric is the extremal Kerr metric. Thus, they were
the first to find a situation in which the matter can approach its own
horizon, now called a quasihorizon. Such systems are precisely quasi-black
holes. Recently, these rotational counterparts of quasi-black holes were
further considered by Meinel \cite{meinel1}, although the term quasi-black
hole was still not coined there (see also \cite{meinel2,meinel3}). Rotating
objects have astrophysical relevance, so it is certainly of interest to
consider the rotating versions of quasi-black holes. For a distant observer,
such rapidly rotating bodies would look almost indistinguishable from black
holes.

The paper of Meinel \cite{meinel1} is an important development of the
subject, and it contains a very strong claim that should be further explored
\cite{meinel1}. On the basis of an analysis of the mass formulas alone,
Meinel \cite{meinel1} argued that the only suitable candidate to the role of
a limiting configuration (i.e., a quasi-black hole, or a body that
approaches its own gravitational radius) corresponds to the extremal case.
So the conclusion made in our work \cite{qbh}, that static quasi-black holes
should be extremal, relying heavily on the properties of the finiteness of
the surface stresses that arise in the quasi-black hole limit, and also
through the mass formula afterward \cite{paper1}, have a seemingly analogous
statement in the rotating stationary extremal case. The conclusion drawn in
\cite{meinel1} was inferred only from the formula for the mass, and
moreover, surface stresses were not taken into account at all. However, with
the know how one can take from the static case \cite{paper1,qbh}, one is led
to consider these stresses in order to be able to make more general
statements. Moreover, we will see that without an appropriate account for
the stresses the analysis would remain essentially incomplete. Thus, bearing
in mind both the theoretical interest of stationary configurations on the
threshold of the formation of a horizon and their potential astrophysical
significance, one should also allow for surface stresses, either finite or
infinite, in the stationary case.

In this paper, we thus consider stationary configurations with surface
stresses which can either be finite or grow without bound when the rotating
quasi-black hole is being formed. In this sense, the statements in \cite
{meinel1} are generalized. We extend further the analysis and consider
configurations not only with mass and angular momentum \cite{meinel1}, as
well as surface stresses, but also with electrical charge. We find the
angular momentum and mass formulas for this general charged stationary
configuration. We also elucidate what makes the extremal state in the
stationary case a distinguished configuration.

As in the static case \cite{paper1}, the two issues, one of the relation
between surface stresses and the mass formula for quasi-black holes, the
other of the mass formulas for quasi-black holes and pure black holes, are
interconnected. We argue there is a close correspondence between the mass
formulas for quasi-black holes and black holes in all cases, non-extremal
and extremal, despite the fact that the physical nature of these objects
(see \cite{qbh}) and the derivation of the mass formula itself are quite
different. This, of course, makes the encountered close relationship between
the formulas non-trivial. Our analysis has also rather unexpected
consequences for the general relativistic counterpart of the classical
Abraham-Lorentz model for electron, connected with the distinguished role
played by the stationary quasi-horizon in the extremal case.

\section{Metric form for rotating stationary configurations}

\subsection{Metric form and the definition of a stationary quasi-black hole}
\label{metricform}

Let us have a distribution of matter in a gravitational field which does not
depend on time. Put the four-dimensional spacetime metric $ds^{2}=g_{\mu \nu
}dx^{\mu }dx^{\nu }$, with $\mu ,\nu $ being spacetime indices, in the form
\begin{equation}
ds^{2}=-N^{2}dt^{2}+g_{ik}\left( dx^{i}+N^{i}dt\right) \left(
dx^{k}+N^{k}dt\right) ,  \label{metricgeneral}
\end{equation}
where, we use $0$ as a time index, and $i,k=1,2,3$ as spatial indices. In
addition, $N$ and $N^{i}$ are the lapse function and shift vector which
depend in general on the spatial coordinates $x^{i}$.

{}From (\ref{metricgeneral}), the metric of a stationary axially-symmetric
system can be written in a useful form by putting $N^3=N^\phi=-\omega$,
where $\phi$ is the azimuthal coordinate and $\omega$ an angular velocity,
and the other $N^i$ obey $N^i=0$. We denote the radial coordinate by $l$ and
put the radial potential $g_{ll}=1$. If we further define a cylindrical
coordinate $z$, the metric can be written in the form,
\begin{equation}
ds^{2}=-N^{2}dt^{2}+dl^{2}+g_{{z}{z}}dz^{2} +g_{\phi \phi }(d\phi -\omega
dt)^{2},  \label{mv}
\end{equation}
an axially symmetric form, where the metric coefficients depend on $l$ and
$
z $ \cite{vis}.

In \cite{paper1} we extended the definition of a quasi-black hole from the
spherically symmetric and extremal case \cite{qbh} to a generic static case.
Now, we extend it further to a stationary spacetime. Several points of
\cite
{paper1} are repeated with the reservation that now $g_{00}\neq -N^{2}$ due
to the terms responsible for rotation. Namely, consider a configuration
depending on a parameter $\varepsilon $ such that (a) for small but non-zero
values of $\varepsilon$ the metric is regular everywhere with a
non-vanishing lapse function $N$, at most the metric contains only
delta-like shells, (b) taking as $\varepsilon $ the maximum value of the
lapse function on the boundary $N_{\mathrm{B}}$, then in the limit $
\varepsilon \rightarrow 0$ one has that the lapse function $N\leq $ $N_{
\mathrm{B}}\rightarrow 0$ everywhere in the inner region, (c) the
Kretschmann scalar $\mathrm{Kr}$ remains finite in the quasihorizon limit.
This latter property implies another important property which can be stated
specifically, namely, (d) the area $A$ of the two-dimensional boundary $l=
\mathrm{const}$ attains a minimum in the limit under consideration, i.e., $
\lim_{\varepsilon \rightarrow 0}\frac{\partial A}{\partial
l}|_{l^{\ast }}=0$, 
where $l^{\ast }$ is the value of $l$ at the quasi-horizon. In addition,
now we also require that in the limit under discussion $\omega \rightarrow
\omega_{\mathrm{h}}=$\textrm{const} everywhere in the inner region. Here $
\omega_{\mathrm{h}}$ corresponds to the angular velocity of a black hole to
which the quasi-black hole metric tends outside. Without this property, the
differential rotation inside would serve to distinguish a black hole and
quasi-black hole metrics and, thus, the definition of a quasi-black hole
would not have physical meaning. The constancy of $\omega_{\mathrm{h}}$ is a
known property of black holes and can be substantiated by the regularity of
the curvature invariants \cite{vis}. It is worth also mentioning that the
system under consideration can in general represent either a compact body
with a well defined junction to a electrovacuum solution, or a dispersed
distribution of matter.

\subsection{Other discussions}

For a situation in which the body's surface approaches the would-be horizon
(quasi-horizon), we take advantage of the asymptotics of the lapse function
$
N$ and the function $\omega $ near the horizon \cite{vis}. Then, for the
non-extremal case, approximating the metric in the outer region by that of a
black hole, we have the following relations,
\begin{equation}
N=\kappa \,l+O(l^{3}),  \label{N}
\end{equation}
and
\begin{equation}
\omega =\omega _{\mathrm{h}}+O(l^{2})\text{,}  \label{om}
\end{equation}
where $\kappa $ is the surface gravity at the horizon obeying $\kappa =
\mathrm{constant\neq 0}$, and $\omega _{\mathrm{h}}$ is the horizon value of
$\omega $ obeying $\omega _{\mathrm{h}}=\mathrm{constant}$. For the extremal
case $\kappa =0$ similarly to the static case and $N\sim \exp (-Bl)$, $B=
\mathrm{constant}$ \cite{paper1}. For the ultraextremal case \cite{paper1},
one has $N\sim l^{-n}$ and $\kappa =0$. In both these two cases we assume
that near the horizon
\begin{equation}
\omega =\omega _{\mathrm{h}}+a_{1}N+a_{2}N^{2}+...  \label{w}
\end{equation}
where $\omega _{\mathrm{h}}$ and $a_{1},\,a_{2},...$ are constants.

Two reservations are in order. First, the relevance of the asymptotics
in Eq. (\ref{w}) in our context should follow from the analysis of the
near-horizon behavior of the scalars (such as Ricci, Kretschmann, and
other scalars), composed out of the curvature components. Such an
analysis was performed in \cite{vis} for the non-extremal case, only
partially for the ultraextremal one, and not at all for the extremal
case. Strictly speaking, the necessity of the asymptotics (\ref{w})
was not proved formally in \cite {vis} for extremal and ultraextremal
horizons and remains a gap to be filled. However, its derivation
represents a formal problem on its own that would take us far
afield. Therefore, we simply assume the validity of the Taylor
expansion given in Eq. (\ref{w}).  Second, Eq.  (\ref{w}) is assumed
to be an expansion with respect to the quasi-horizon for the outer
region. On the other hand, we assume (as explained at the end of
Sec. \ref{metricform}) that $\omega \rightarrow \omega
_{\mathrm{h}}={\rm const}$ everywhere in the inner region. As a result,
there is a jump of the normal derivative $\frac{\partial \omega
}{\partial l}$ in the quasi-horizon limit for the non-extremal
case. This is similar to what happens to the lapse function
\cite{paper1}.

\section{The angular momentum and mass formulas for the stationary
case: a stationary axially-symmetric rotating configuration spacetime}

If the matter is joined onto a vacuum spacetime then one has to be
careful and use the junction condition formalism \cite{isr,mtw}. The
angular momentum and mass of the matter distribution can be written as
integrals over the region occupied by matter and fields. Defining
$T^\nu_\mu$ as the stress-energy tensor, the momentum $J_i$ relative
to a coordinate $x^i$ is given by
\begin{equation}
J_i=-\int T_i^{0} \sqrt{-g} \,d^{3}x\,,  \label{Jtot}
\end{equation}
where $g$ is the determinant of the metric $g_{\mu\nu}$. When the coordinate
$x^i$ is angular and cyclic $\phi$ say then $J_\phi$ is an angular momentum
and one puts $J_\phi\equiv J$ (see, e.g., \cite{p}). The mass of the matter
distribution can be written as an integral over the region occupied by
matter and fields. It is given by the Tolman formula \cite{tolman} (see also
\cite{ll} and \cite{p}),
\begin{equation}
M=\int (-T_{0}^{0}+T_{k}^{k})\sqrt{-g}\,d^{3}x\,.  \label{mtot}
\end{equation}
This is the starting point of our analysis. We discuss these integrals for
an axially-symmetric rotating matter distribution in a an axially-symmetric
rotating spacetime. In summary, we consider the stationary case,
generalizing the static case discussed in a previous paper \cite{paper1}.
For the angular momentum and mass formulas for black holes, rather than
quasi-black holes, see \cite{bch,cart73,bard73,sm}, and, particularly \cite
{fn} for the angular momentum formula.

\subsection{The various angular momenta and masses}

We consider the stationary case, with axial symmetry. We assume that the
body has a well-defined quasi-black hole limit.

\subsubsection{Total angular momentum, and total mass}

Let us have a distribution of matter and a gravitational field which do not
depend on time. Note also from Eq. (\ref{mv}) that $\sqrt{-g}=N\sqrt{g_3}$,
where $g_3$ is the determinant of the spatial part of the metric (\ref{mv}),
i.e., is the determinant of the metric on the hypersurface $t=\mathrm{
constant}$. We consider first the angular momentum. Then from Eq.
(\ref{Jtot}), the total angular momentum $J$ is given by
\begin{equation}
J=-\int T_{\phi }^{0}\,N\sqrt{g_3}\,d^{3}x\,.  \label{Jstationary}
\end{equation}
Then, the total value of the angular momentum (\ref{Jstationary}) can be
split into three contributions the inner, the surface, and the outer, such
that,
\begin{equation}
J_{\mathrm{tot}}=J_{\mathrm{in}}+J_{\mathrm{surf}}+ J_{\mathrm{out}}\,.
\label{jtotal}
\end{equation}
Next, we consider the mass, which can be written as an integral over the
region occupied by matter and fields,
\begin{equation}
M=\int (-T_{0}^{0}+T_{k}^{k})\,N\sqrt{g_3}\,d^{3}x\,.
\label{mtotstationary}
\end{equation}
{}From Eq. (\ref{mtotstationary}) it is again convenient here to compose the
linear split of the total mass into three different contributions, the
inner, the surface mass, and the outer masses, such that
\begin{equation}
M_{\mathrm{tot}}=M_{\mathrm{in}}+M_{\mathrm{surf}}+ M_{\mathrm{out}}\,.
\label{split}
\end{equation}
Note that for the outer mass a long-range electromagnetic field may be
present.

\subsubsection{Inner angular momentum and mass}

As in the static case \cite{paper1}, one has for a quasi-black hole that
$N_{
\mathrm{B}}\rightarrow 0$, where $N_{\mathrm{B}}$ is the value of $N$ at the
boundary as well as $N\rightarrow 0$ for the whole inner region. So, the
inner contribution to the angular momentum vanishes in the quasi-black hole
limit due to the factor $N$, i.e.,
\begin{equation}
J_{\mathrm{in}}= 0\,.  \label{jint}
\end{equation}
For the same reasons, and analogously to the static case, 
the inner contribution to the mass vanishes,
\begin{equation}
M_{\mathrm{in}}=0\,.  \label{min}
\end{equation}

\subsubsection{Surface angular momentum and mass}

Now consider the contribution of the surface to the angular momentum and
mass. First, the angular momentum. One has,
\begin{equation}
J_{\mathrm{surf}}=-\int_{\mathrm{surface}} T_{\phi }^{0}\,N\sqrt{g_3}
\,d^{3}x\,.  \label{Jstationarysurface}
\end{equation}
Defining $\gamma$ as
\begin{equation}
\gamma =-\frac{1}{2N^{2}}g_{\phi \phi } \frac{\partial \omega }{\partial l}
\,,  \label{gamma}
\end{equation}
we can put Eq. (\ref{Jstationarysurface}) in the form
\begin{equation}
J_{\mathrm{surf}}=\frac{1}{8\pi}\int \gamma \,N\, d\sigma\,,  \label{j}
\end{equation}
where $d\sigma$ is the two-dimensional surface spanned by
$t=\mathrm{constant
}$, $l=\mathrm{constant}$. Now, for the pure black hole case, the angular
momentum of the horizon is equal to \cite{fn}
\begin{equation}
J_{\mathrm{h}}=-\frac{1}{8\pi} \int_{\mathrm{horizon}}\xi^{\mu\,;\,\nu}\,d
\sigma _{\mu \nu }\,,  \label{jh}
\end{equation}
where the integration is taken over the horizon surface with element $
d\sigma _{\mu \nu}$, and $\xi^\mu$ are the components of the rotational
Killing vector $\xi$, which is given by $\xi=\frac{\partial }{\partial
\phi}$, 
and a semi-colon denotes covariant derivative (see, e.g., \cite{fn}). One
can now show that in the quasi-black hole limit, Eq. (\ref{j}) reduces to
Eq. (\ref{jh}). Indeed, taking a cross section of the metric (\ref{mv}) such
that $t=\mathrm{constant}$ and $l=\mathrm{constant}$, and developing
expression (\ref{jh}) explicitly, one finds that in the quasi-black hole
limit (\ref{j}) coincides exactly with (\ref{jh}), so that
\begin{equation}
J_{\mathrm{surf}}=J_{\mathrm{h}}\,,  \label{jequal}
\end{equation}
where $J_{\mathrm{h}}$ should now be interpreted as the angular momentum of
the quasi-black hole. For the non-extremal case it is finite and, in
general, non-zero. For the extremal case it is also finite and in general
non-zero, as it follows from (\ref{w}) and from $N\sim \exp (-Bl)$ as $
l\rightarrow \infty$. Only in some special extremal configurations the
surface stresses vanish (see, e.g., the example of the spherically symmetric
static system composed of extremal charged dust (see \cite{bon-d} and
references therein)). For the ultraextremal case, defined above, assuming
the validity of the asymptotic expansion (\ref{w}) one finds that the
surface contribution to the angular momentum vanishes.

Now consider the contribution of the surface to the mass,
\begin{equation}
M_{\mathrm{surf}}=\int_{\mathrm{surface}}(-T_{0}^{0}+T_{k}^{k})\,N\sqrt{g_{3
}
}\,d^{3}x\,\,.
\end{equation}
As in the static case there are delta-like contributions, given by
\begin{equation}
S_{\mu }^{\nu }=\int T_{\mu }^{\nu }\,dl\,,
\end{equation}
where the integral is taken across the shell. Define $\alpha $ as,
\begin{equation}
\alpha =8\pi (S_{a}^{a}-S_{0}^{0})\,.
\end{equation}
Then, from a combination of the equations above, we get,
\begin{equation}
M_{\mathrm{surf}}=\frac{1}{8\pi }\int \alpha \,N\,d\sigma \,,  \label{mrot}
\end{equation}
where $d\sigma $ is the surface element. Now, one also has the relationship
$
8\pi S_{\mu }^{\nu }=[[K_{\mu }^{\nu }]]-\delta _{\mu }^{\nu }[[K]],\label
{s1}$ where $K_{\mu }^{\nu }$ is the extrinsic curvature tensor, $
[[...]]=[(...)_{+}-(...)_{-}]$, subscripts \textquotedblleft +" and
\textquotedblleft -" refer to the outer and inner sides, respectively (see,
e.e.g, \cite{isr,mtw}). Also, $K_{\mu \nu }=-n_{\mu ;\nu }$, where at the
boundary surface $N=\mathrm{constant}$, and the normal unit vector is $
n_{\mu }\sim N_{;\mu }$. Thus, $\alpha =-[[2K_{0}^{0}]]$, and further
calculations give
\begin{equation}
\alpha =\frac{2}{N}\left[ \left( \frac{\partial N}{\partial l}\right)
_{+}-\left( \frac{\partial N}{\partial l}\right) _{-}\right]
+\frac{1}{N^{2}}
\,g_{\phi \phi }\,(\omega -\omega _{\mathrm{h}})\,\frac{\partial \omega }{
\partial l}\,,  \label{b}
\end{equation}
and so,
\begin{equation}
M_{\mathrm{surf}}=\frac{1}{4\pi }\int_{\mathrm{surf}}\left[ \left[ \left(
\frac{\partial N}{\partial l}\right) _{+}-\left( \frac{\partial N}{\partial
l
}\right) _{-}\right] +\frac{2}{N}\,g_{\phi \phi }\,\omega \,\frac{\partial
\omega }{\partial l}\right] d\sigma \,.  \label{mrot2}
\end{equation}
Now, as our surface approaches the would-be horizon, i.e., the
quasi-horizon, we take advantage of the asymptotics near the horizon of the
lapse function $N$ and of the function $\omega $. Thus, taking into account
expression (\ref{j}) and the asymptotics (\ref{N}) and (\ref{om}) in the
non-extremal case, or Eq. (\ref{w}) in the extremal or ultraextremal cases,
we obtain
\begin{equation}
M_{\mathrm{surf}}=\frac{\kappa A_{\mathrm{h}}}{4\pi }+2\,\omega
_{\mathrm{h}
}J_{\mathrm{h}}\,,  \label{mh}
\end{equation}
where $\kappa $ is the surface gravity of the quasi-black hole. So, in
relation to the contribution of the surface stresses to the mass, what was
said in the static case \cite{paper1} applies here to the first term of Eq.
(\ref{mh}). Namely, in the non-extremal case the stresses are infinite but
their contribution is finite and non-zero, in the extremal case they are
finite but their contribution vanishes, and in the ultraextremal case the
stresses themselves are zero, so the contribution to the mass is zero as
well. As far as the second, new, term in (\ref{mh}) is concerned, it follows
that the surface contribution is non-zero for the non-extremal and extremal
cases but vanishes in the ultraextremal one. Note also, that although for
the non-extremal ($\kappa \neq 0$) case on one hand and for the extremal and
ultraextremal ($\kappa =0$) cases on the other, we have used different
asymptotics of the metric coefficients near the quasihorizon, the smooth
limiting transition $\kappa _{\mathrm{h}}\rightarrow 0$ can be made in the
formula (\ref{mh}) for the surface contribution as a whole, surely. Since
Eq. (\ref{mh}) shows clearly that one cannot ignore surface stresses
contribution in the non-extremal case, the analysis in \cite{meinel1} is
incomplete. It omits from the very beginning just the most important feature
of non-extremal configurations in their confrontation with the extremal
ones. This means the final conclusion of \cite{meinel1} hangs in mid-air.
That is, one could na\"{\i}vely think that one could simply restrict oneself
to the case of vanishing stresses but in the problem under discussion this
is impossible. Indeed, we have just seen that the stresses enter the mass
formulas via the quantity $\alpha $, so in the case of vanishing stresses $
M_{\mathrm{surf}}$ would also vanish. But this does not happen.

\subsubsection{Outer angular momentum and mass}

The outer angular momentum is given generically by the expression,
\begin{equation}
J_{\mathrm{out}}=-\int_{\mathrm{outer}}
T_{\phi }^{0}\,N\sqrt{g_3}\,d^{3}x\,.
\label{Jstationaryouter}
\end{equation}
The outer mass is given generical by the expression
\begin{equation}
M_{\mathrm{out}}=\int_{\mathrm{outer}}(-T_{0}^{0}+T_{k}^{k}) \,N \sqrt{g_3}
\,d^{3}x\,.
\end{equation}
Further, we may split $M_{\mathrm{out}}$ into an electromagnetic part
$M_{\,
\mathrm{out}}^{\mathrm{em}}$, and a non-electromagnetic part,
$M_{\;\mathrm{
out}}^{\mathrm{matter}}$ say, for the case of dirty black holes or dirty
quasi-black holes, exactly in the manner as it was already done in \cite
{cart73}, and obtain
$M_{\mathrm{out}}=M_{\,\mathrm{out}}^{\mathrm{em}}+M_{\;
\mathrm{out}}^{\mathrm{matter}}$. Since $M_{\,\mathrm{out}}^{\mathrm{em}
}=\varphi _{\mathrm{h}}Q$ (see \cite{paper1} for details), where $\varphi_{
\mathrm{h}}$ is the electric potential on the horizon in the case of black
holes, and the electric potential on the quasihorizon in the case of
quasi-black holes, and $Q$ is the corresponding electric charge, one finds
\begin{equation}
M_{\mathrm{out}}=\varphi _{\mathrm{h}}Q+M_{\;\mathrm{out}}^{\mathrm{matter}
}\,.  \label{outerm2}
\end{equation}

\subsection{The angular momentum and mass formulas}


Putting all together, for the quasi-black hole case, and recalling that
$J_{
\mathrm{in}}$ goes to zero, the total angular momentum is equal to
\begin{equation}
J=J_{\mathrm{h}}+J_{\mathrm{out}}\,.
\end{equation}
In vacuum, if matter is absent or negligible outside, we have only $J_{
\mathrm{h}}$ i.e., the total angular momentum is the quasi-black hole
angular momentum.

In a similar way, recalling that ${M}_{\mathrm{in}}$ goes to zero, we find
the total mass is equal to
\begin{equation}
M=\frac{\kappa A_{\mathrm{h}} }{4\pi }++2\omega _{\mathrm{h}}J_{\mathrm{h}
}+\varphi_{\mathrm{h}}Q +M_{\;\mathrm{out}}^{\mathrm{matter}}\,.
\label{tot}
\end{equation}
Equation (\ref{tot}) is the mass formula for stationary quasi-black holes.
But on closer inspection it is nothing else than the mass formulas for black
holes \cite{bch,cart73,bard73,sm,fn}. Note that for the extremal case, the
term $\frac{\kappa A_{\mathrm{h}}}{4\pi }$ in Eq. (\ref{tot}) goes to zero,
since $\kappa$ is zero. In vacuum, if matter is absent or negligible
outside, we return to
\begin{equation}
M_{\mathrm{h}}=\frac{\kappa A_{\mathrm{h}}}{4\pi }+ 2\omega
_{\mathrm{h}}J_{
\mathrm{h}} +\varphi _{\mathrm{h}}Q\,,  \label{totsmarr}
\end{equation}
which is Smarr's formula, but now for quasi-black holes, i.e., formula
(\ref{totsmarr}) is equal to a formula first found 
by Smarr for Kerr-Newman black
holes \cite{sm}. Here we see it holds good for rotating quasi-black holes as
well. It is also worth noting that in our approach we did not restrict
ourselves to a compact body rotating with a constant angular velocity in
vacuum as it was done in \cite{meinel1}. Instead, we have admitted all types
of rotation, including differential and rigid rotations, as well as matter
distribution outside the quasihorizon. Now, in the context of the uniqueness
theorems, it is specially interesting to trace how the configuration of a
self-gravitating rotating body approaches an outer vacuum Kerr-Newman
metric. In this context, by allowing infinite surface stresses we can
conjecture, resorting to the uniqueness theorems (see, e.g., \cite{fn}) and
continuity arguments between a horizon and a quasihorizon with outer vacua,
that the generic Kerr-Newman metric, and so the Kerr metric, is an outer
metric for some type of matter that allows infinite stresses. In addition,
Eqs. (\ref{tot})-(\ref{totsmarr}) reduce to the static case considered in
\cite{paper1} for $\omega_{\mathrm{h}}=0$.

Consider, as an example, the case where there is only rotation and no
electrical field nor matter in the outer region. Thus, the exterior to
the quasi-black hole is described by the extremal Kerr metric. Then
$\kappa =0$, and $M_{\mathrm{h}}=2\omega
_{\mathrm{h}}J_{\mathrm{h}}$. Also see \cite {paper1} for the case
$\omega _{\mathrm{h}}=0$ and the example for the charged static case.

Thus, we have traced how the total mass of a quasi-black hole, which
can be defined at asymptotical infinity as usual \cite{Arn,Y}, is
distributed among different terms including the contribution from the
quasihorizon. We have found perfect correspondence with the black hole
case.

\section{Conclusions}

There are three main topics and conclusions that can be taken out of our
results: (i) With rotation and charge the problem of a self-consistent
analog of an elementary particle in general relativity is much more
interesting than without rotation. If one wants a classical model for the
electron one certainly should look for including rotation, see \cite{paper1}
and \cite{vf,ep} for the static case (see also \cite{Arn,Y}). As a
by-product, we have obtained that an extremal quasi-black hole can serve as
a classical model of an Abraham-Lorentz electron in that both the inner and
surface contribution of non-electromagnetic forces vanish. In doing so, we
showed that one may weaken the requirement of vanishing surface stresses
since the finite stresses have zero contribution to the total mass. (ii)
Here we have traced how the limiting transition from a stationary
configuration to the quasi-black hole state reveals itself in the mass
formula, going thus beyond the static case \cite{paper1} and beyond what was
found in \cite{meinel1} for a particular set of stationary configurations
(see also \cite{meinel2,meinel3}). It turns out that the perfect one-to-one
correspondence between the different contributions for the total mass of a
quasi-black hole and the mass formula for black holes persists in the
generic stationary case. In particular, the inner contribution to the total
mass vanishes in the quasi-black hole limit (it is absent in the black hole
case from the very beginning). The contribution of the surface stresses
corresponds just to the contribution from the horizon surface of a black
hole. This is not trivial, since the corresponding terms have quite
different origins. In the quasi-black hole case they are due to the boundary
between both sides of the surface. Meanwhile, in the black hole case only
one (external) side is relevant and the integrand over this surface has
nothing to do with the expression for surface stresses. Nonetheless, both
terms coincide in the limit under discussion. Similar results were obtained
for the angular momentum of the rotating configurations. As bodies with
rotation occur widely in nature, the results obtained may have astrophysical
implications. (iii) The difference between non-extremal and extremal
quasi-black holes consists in that in the first case the surface stresses
give finite contribution to the total mass, but become infinite, while in
the second case they give zero contribution to the total mass, but are
finite. As far as the mass is concerned, in the non-extremal case the
surface of a quasi-black hole appears in a way similar to a membrane in the
membrane paradigm setup \cite{membr}, whereas in the extremal one we have in
general a ``membrane without membrane'' \cite{wheeler}. The system with
infinite stresses was rejected in \cite{qbh}, since it looks unphysical, and
thus in \cite{qbh} only extremal black holes were considered. However,
consideration of such systems helps in understanding better the relationship
between quasi-black holes and black holes and the distinction between
non-extremal and extremal limits. With its astrophysical as well as
theoretical importance, the rotating case, as we was disucssed here,
acquires added relevance.

\begin{acknowledgments}
O. Z. thanks Centro Multidisciplinar de Astrof\'{\i}sica -- CENTRA for
hospitality. This work was partially funded by Funda\c c\~ao para a
Ci\^encia e Tecnologia (FCT) - Portugal, through project
PPCDT/FIS/57552/2004.
\end{acknowledgments}

\end{document}